# Reply to "Comment on "Fully covariant radiation force on a polarizable particle""

A. A. Kyasov and G. V. Dedkov

Nanoscale Physics Group, Kabardino—Balkarian State University, Nalchik, Russia

**Abstract.** We reply to comment on recent paper by Pieplow and Henkel (New J. Phys. 15 (2013) 023027) and our results made by Volokitin and Persson (arXiv: 1405.2525)

Recent work [1] summed a certain result of years discussion in the problem of fluctuation electromagnetic interaction (FEI) of a small moving particle with a flat surface (configuration 1) and a homogeneous radiation background. The authors of [1] have constructed a fully covariant theory which made it possible to confirm in detail previously obtained results [2-4] relating to the forces acting on a particle moving with an arbitrary relativistic velocity near a smooth dielectric surface and in homogeneous radiation background (photonic gas).

As a result, the main characteristics of FEI in configuration 1, namely conservative-dissipative forces (normal and parallel to the surface and the particle velocity) and the rate of particle heating (cooling) obtained in a series of our works [2-4] ceased to cause any objections in the scientific literature. As to another configuration of two flat surfaces in relative motion (configuration 2), a more or less satisfactory agreement between the results of different authors in this problem is still absent to date [5-9].

However, quite recently, the results of our works [2-4] and [1] have raised the doubts in a paper by Volokitin and Persson (VP) once again [10]. Before analyzing this work, it is worth noting that for a long period of time [11, 5] VP were considering the problem of dissipative tangential force in configuration 2 based on the dynamic generalization of the Lifshitz theory [12], which holds in the case of total thermal equilibrium in the system of two thick plates at rest. In [6], VP have developed a relativistic variant of this theory in the case of thermal and dynamical disequilibrium. However, when they attempted to pass from configuration 2 to configuration 1 using the limiting transition to a rarified medium for one of the plates, all the results obtained for the characteristics of FEI in configuration 1 turned out to be wrong. A detailed criticism of these results [5,6,11] is presented in our works [13]. Incorrect results in configuration 1 indicate incorrectness of the results in configuration 2 --the basic configuration used by VP.

This time, Volokitin and Persson made a next attempt to obtain the characteristics of FEI in configuration 1, using the corresponding results for configuration 2. Moreover, by badly understandable reasons they took into consideration only the contribution from evanescent surface modes ($k > \omega/c$). In our designations, the final results presented by VP (Eqs. (25), (35) and (33) in [10]) for the components of forces $F_x, F_z$ (tangential and normal to the surface) and the rate of particle heating $\dot{Q}$ have the form

$$F_x^{(PV)} = -\frac{\hbar \gamma^2}{2\pi^2} \int_0^\infty d\omega \int_{k>\omega/c} d^2k\, k_x \frac{\exp(-2q_0 z_0)}{q_0} [\chi_e(\omega,\mathbf{k})\Delta_e''(\omega) + \chi_m(\omega,\mathbf{k})\Delta_m''(\omega)] \alpha''(\gamma\omega^+) \cdot$$
$$\cdot \left[ \coth\left(\frac{\hbar\omega}{2k_B T_2}\right) - \coth\left(\frac{\gamma\hbar\omega^+}{2k_B T_1}\right) \right] , \quad (1)$$

$$F_z^{(PV)} = -\frac{\hbar \gamma^2}{2\pi^2} \int_0^\infty d\omega \int_{k>\omega/c} d^2k\, \exp(-2q_0 z_0) \cdot$$
$$\cdot \left\{ \begin{array}{l} [\chi_e(\omega,\mathbf{k})\Delta_e'(\omega) + \chi_m(\omega,\mathbf{k})\Delta_m'(\omega)] \alpha''(\gamma\omega^+) \coth\left(\frac{\gamma\hbar\omega^+}{2k_B T_1}\right) + \\ [\chi_e(\omega,\mathbf{k})\Delta_e''(\omega) + \chi_m(\omega,\mathbf{k})\Delta_m''(\omega)] \alpha'(\gamma\omega^+) \coth\left(\frac{\hbar\omega}{2k_B T_2}\right) \end{array} \right\} , \quad (2)$$

$$\dot{Q}^{(PV)} = \frac{\hbar \gamma^2}{2\pi^2} \int_0^\infty d\omega \int_{k>\omega/c} d^2k\, \omega \frac{\exp(-2q_0 z_0)}{q_0} [\chi_e(\omega,\mathbf{k})\Delta_e''(\omega) + \chi_m(\omega,\mathbf{k})\Delta_m''(\omega)] \alpha''(\gamma\omega^+) \cdot$$
$$\cdot \left[ \coth\left(\frac{\hbar\omega}{2k_B T_2}\right) - \coth\left(\frac{\gamma\hbar\omega^+}{2k_B T_1}\right) \right] , \quad (3)$$

where $T_1$ is the particle temperature, $T_2$ is the temperature of a thick plate (half-space $z<0$) having the dielectric and magnetic permeabilities $\varepsilon(\omega), \mu(\omega)$; $z_0$ is the particle distance from the plate, $\gamma = (1-\beta^2)^{-1/2}$ is the relativistic factor, $\beta = V/c$, $\omega^+ = \omega + k_x V$,

$$\chi_e(\omega,\mathbf{k}) = 2(k^2 - k_x^2 \beta^2)\left(1 - \frac{\omega^2}{k^2 c^2}\right) + \left(\frac{\omega + k_x V}{c}\right)^2 , \quad (4)$$

$$\chi_m(\omega,\mathbf{k}) = 2k_y^2 \beta^2 \left(1 - \frac{\omega^2}{k^2 c^2}\right) + \left(\frac{\omega + k_x V}{c}\right)^2 , \quad (5)$$

$$\Delta_e(\omega) = \frac{\varepsilon(\omega)q_0 - q}{\varepsilon(\omega)q_0 + q}, \Delta_m(\omega) = \frac{\mu(\omega)q_0 - q}{\mu(\omega)q_0 + q} \quad . \tag{6}$$

Let us compare Eqs. (1)—(3) with the corresponding results in our works (for example, [2,3])

$$F_x^{(DK)} = -\frac{\hbar\gamma}{2\pi^2}\int_0^\infty d\omega \int_{k>\omega/c} d^2k\, k_x \frac{\exp(-2q_0 z_0)}{q_0}[\chi_e(\omega,\mathbf{k})\Delta_e''(\omega) + \chi_m(\omega,\mathbf{k})\Delta_m''(\omega)]\alpha''(\gamma\omega^+) \cdot$$
$$\cdot\left[\coth\left(\frac{\hbar\omega}{2k_B T_2}\right) - \coth\left(\frac{\gamma\hbar\omega^+}{2k_B T_1}\right)\right] \tag{7}$$

$$F_z^{(DK)} = -\frac{\hbar\gamma}{2\pi^2}\int_0^\infty d\omega \int_{k>\omega/c} d^2k\, \exp(-2q_0 z_0) \cdot$$
$$\cdot\left\{\begin{array}{l}[\chi_e(\omega,\mathbf{k})\Delta_e'(\omega) + \chi_m(\omega,\mathbf{k})\Delta_m'(\omega)]\alpha''(\gamma\omega^+)\coth\left(\frac{\gamma\hbar\omega^+}{2k_B T_1}\right) + \\ [\chi_e(\omega,\mathbf{k})\Delta_e''(\omega) + \chi_m(\omega,\mathbf{k})\Delta_m''(\omega)]\alpha'(\gamma\omega^+)\coth\left(\frac{\hbar\omega}{2k_B T_2}\right) \end{array}\right\} \tag{8}$$

$$\dot{Q} = \frac{\hbar\gamma}{2\pi^2}\int_0^\infty d\omega \int_{k>\omega/c} d^2k\, (\omega + k_x V)\frac{\exp(-2q_0 z_0)}{q_0}[\chi_e(\omega,\mathbf{k})\Delta_e''(\omega) + \chi_m(\omega,\mathbf{k})\Delta_m''(\omega)]\alpha''(\gamma\omega^+) \cdot$$
$$\cdot\left[\coth\left(\frac{\hbar\omega}{2k_B T_2}\right) - \coth\left(\frac{\gamma\hbar\omega^+}{2k_B T_1}\right)\right] \tag{9}$$

Comparing (1),(2) with (7),(8) one can see that

$$F_x^{(PV)} = \gamma F_x^{(DK)}$$
$$F_z^{(PV)} = \gamma F_z^{(DK)}$$

Therefore, the difference between these results is considerable in the limit $\gamma \gg 1$. Still greater difference is observed between Eqs. (3) and (9), since the integrand expression in (9) contains the shifted frequency $\omega^+ = \omega + k_x V$ instead of $\omega$ before the exponential factor in (3). In this regard, Eqs. (3) and (9) do not coincide even in the nonrelativistic limit. The last error is repeated since the earlier work by VP [14]. Moreover, the formula for the tangential force on a particle moving in photonic gas obtained in [6] proves to be basically wrong since it does not contain the contribution depending on the particle temperature. Contrary to the author's claim [10], our results [2-4, 13] are in complete agreement with [1] both in configuration 1 and in the case of particle motion in the black-body radiation.

Thus, compared to their previous works [5,6,11,14] Volokitin and Persson managed to come a little closer but not fully achieve the right results, which have been obtained 12 years ago [2,3]. In this regard, we claim that recent attempt [10] of obtaining relativistic generalization of the Lifshitz-Pitaevskii theory is untenable. Unfortunately, the authors do not comment their earlier papers, in which they have obtained the results which differ considerably from (1) - (3) and contain fatal errors in the expressions of the forces involving propagating surface modes while continuing to refer to these works.